\documentclass[10pt,conference]{IEEEtran}
\IEEEoverridecommandlockouts

\usepackage{multirow}
\usepackage{url}
\usepackage{mdframed}
\usepackage[normalem]{ulem}
\usepackage{cite}
\usepackage{amsmath,amssymb,amsfonts}
\usepackage{algorithmic}
\usepackage{graphicx}
\usepackage{textcomp}
\usepackage{xcolor}
\usepackage{soul}
\def\BibTeX{{\rm B\kern-.05em{\sc i\kern-.025em b}\kern-.08em
    T\kern-.1667em\lower.7ex\hbox{E}\kern-.125emX}}
    
\newcommand{\university}{\texttt{blinded-university}}
\begin{document}

\title{Strategies, Benefits and Challenges of \\
App Store-inspired Requirements Elicitation
}

\author{\IEEEauthorblockN{Alessio Ferrari}
\IEEEauthorblockA{\textit{ISTI-CNR} \\
alessio.ferrari@isti.cnr.it}
\and
\IEEEauthorblockN{Paola Spoletini}
\IEEEauthorblockA{\textit{Kennesaw State University}, USA \\
pspoleti@kennesaw.edu}
}

\maketitle

\begin{abstract}
App store-inspired elicitation is the practice of exploring competitors' apps, to get inspiration for requirements. This activity is common among developers, but little insight is available on its practical use, advantages and possible issues. This paper aims to study strategies, benefits, and challenges of app store-inspired elicitation, and to compare this technique with the more traditional requirements elicitation interviews. We conduct an experimental simulation with 58 analysts, and collect qualitative data. Our results show that: (1) specific guidelines and  procedures are required to better conduct app store-inspired elicitation; (2) current search features made available by app stores are not suitable for this practice, and more tool support is required to help analysts in the retrieval and  evaluation of competing products; (3) while interviews focus on the \textit{why} dimension of requirements engineering (i.e., goals), app store-inspired elicitation focuses on  \textit{how} (i.e., solutions), offering indications for implementation and improved  usability. 
Our study provides a framework for researchers to address existing challenges and suggests possible benefits to foster app store-inspired elicitation among practitioners. 
\end{abstract}



\maketitle

%
\IEEEpeerreviewmaketitle

\section{Introduction}
\label{sec:intro}
Requirements can be elicited from stakeholders through several techniques, including  interviews, focus groups, workshops, and questionnaires~\cite{dieste2010systematic}. In the last decade, together with the increasing growth of the market of mobile applications (\textit{apps}), app stores and user reviews offered a new means to gather requirements directly from users~\cite{genc2017systematic,martin2016survey}. Automated solutions have been developed to 
classify reviews~\cite{maalej2016automatic}, mining non-functional requirements~\cite{jha2019mining},  trace reviews with other  artifacts~\cite{palomba2017recommending,noei2019towards,haering2021automatically,oehri2020same}, and other tasks~\cite{martin2016survey,dkabrowski2022analysing}. 
Information from app stores is also used by developers in their daily practice, to prioritise features based on user feedback~\cite{al2019app}, and even to look into competing apps to better understand and exploit the existing market. In particular, the recent survey by Al-Subaihin \textit{et al.}~\cite{al2019app}, involving 186 developers from 36 countries, shows that \textit{The majority of surveyed developers use it [the app store] to explore apps related to their application domain to gain an understanding of the expected user experience and anticipate features}. In this paper, we refer to this practice with the name \textit{app store-inspired elicitation} (ASE). To support this activity, researchers have recently developed specific tools to extract  features from competing products, and enable comparison and feature recommendation~\cite{jiang2019recommending,liu2022mining,dalpiaz2019re}. However, despite the common use of ASE, limited information is available on its use in practice, and in particular on what are the specific strategies adopted by requirements analysts and developers to select inspirational apps. 
Furthermore, knowing what are the advantages and difficulties of ASE, also with respect to other elicitation techniques, could help to better understand when and how to use it. 

In this paper, we study the strategies, benefits and challenges of ASE, and we compare this technique to the more traditional \textit{interview-based elicitation} (IBE). 
To this end, we perform an \textit{experimental simulation} involving 58 analysts. These perform a first classical elicitation process with IBE, followed by ASE. The analysts report their rationale for choosing inspirational apps as well as reflections on the different process phases. The qualitative data are analysed to produce lists of themes, with associated codebooks. 

According to the identified themes, different strategies are used for the selection of inspirational apps. These can be driven by well-planned searches based on possibly required features, but also lazily based on what is returned by search engines, according to simple queries concerning the app domain.  
The main emerging benefits concern the possibility of performing \textit{hands-on} evaluation of competitors' products, thus informing the implementation. Challenges include the difficulty of comparing products in the app store, the uninformative content of app reviews, the risk to mimic other apps, and the absence of a structured process to support ASE. The identified core difference with interviews is that IBE focuses on goals, and ASE focuses on solutions. While the former helps to better interpret needs and foster stakeholders' relationship, the latter supports  feasibility assessment,  prevention of usability issues, and generalisation of an app for a wider market.

Our results can be useful to researchers, as our list of strategies and challenges provide \textit{motivations} to develop further  knowledge, techniques and tools around ASE. In addition, the identified benefits can help practitioners to leverage ASE and combine it with more traditional elicitation practices.

\section{Related Work}
\label{sec:related}
Software engineering supported by app store mining is a widely studied topic. The survey by Martin \textit{et al.}~\cite{martin2016survey} gives a comprehensive overview of the different tasks considered in the literature, while the systematic reviews by Genc-Nayebi and Abran ~\cite{genc2017systematic}, and by Dabrowksi \textit{et al.}~\cite{dkabrowski2022analysing} provide insights on app review mining. 
Our work focuses on \textit{requirements elicitation} using the app store, and in the following, we briefly point to relevant papers in this field.   

The majority of the studies aim to support requirements elicitation by filtering user feedback, typically in the form of app reviews, e.g., by automatically classifying bug reports and feature requests~\cite{maalej2016automatic,jha2018using} or by clustering the feedback to support release planning~\cite{villarroel2016release,scalabrino2017listening}, and extraction of non-functional requirements~\cite{jha2019mining}. More recently, given the insufficiency of the app store in fully supporting app development~\cite{nayebi2018app}, studies have focused on linking app reviews with other requirements-relevant artifacts, such as issue tracking systems~\cite{noei2019towards}, bug reports~\cite{haering2021automatically}, and  tweets~\cite{oehri2020same}. These works leverage user feedback as a primary data source. 

More closely related works to ours are those leveraging \textit{app descriptions} of competing products, besides reviews, to exploit the app market. For example, Jiang \textit{et al.}~\cite{jiang2019recommending} automatically recommend new features, by extracting existing functionalities from similar product descriptions and API names. Similarly, Liu \textit{et al.}\cite{liu2022mining} extract features from competing apps to facilitate app comparison, while Dalpiaz and Parente~\cite{dalpiaz2019re} use reviews for the same goal. Existing tools also support feature extraction from app descriptions. These include  SAFE~\cite{johann2017safe}, and the solution by Harman \textit{et al.}~\cite{harman2012app}, also extended in later studies to find the correlation between features and ratings~\cite{sarro2018customer,finkelstein2017investigating}. The goal of product comparison and feature recommendation was also addressed in an earlier work by Dumitru \textit{et al.}~\cite{dumitru2011demand}, but considering software descriptions from \textit{Softpedia.com}. In the field of software product line engineering, researchers have addressed the problem of market analysis with similar approaches~\cite{bakar2015feature,ferrari2013mining}.  

Besides studies on the \textit{automation} of ASE, some empirical works exist which aim to investigate software engineering issues in mobile app development by means of surveys with practitioners~\cite{holzer2011mobile,joorabchi2013real,francese2017mobile,nayebi2016release}. Among them, the only one that explicitly studies the role of app stores in requirements elicitation is the one by Al-Subaihin \textit{et al.}~\cite{al2019app}. This highlights that ASE is a common practice, but it does not give indications on how this activity is performed or its associated issues. 

\textit{Contribution.} With respect to related works, this is the first one that: (1) provides a list of strategies, benefits, and challenges of ASE and compares it to IBE; (2) deeply investigates ASE in practice, instead of focusing on its automation, or its general adoption by practitioners. 
\section{Study Design}
\label{sec:method}


The proposed study can be classified as an \textit{experimental simulation}, in which we want to evaluate the natural behaviour of analysts in a contrived settings~\cite{stol18}, as done in other studies about  interviews~\cite{debnath21,pitts2007improving,bano2019teaching,ferrari2020sapeer}. This allows us to make a more in-depth analysis with respect to surveys, and to consider human aspects, which have a limited role in tool proposals. 


\subsection{Study Participants}
As requirements analysts, we recruited 58 graduate students enrolled in the first or second semester of the Master in Software Engineering at \university{}.  At the time of the experiment, they were all taking a course on Requirements Engineering, in which they have been introduced to elicitation techniques and user stories.  
80\% of the students have already some professional experience. In particular, 48.33\% have \textit{intense} professional experience, i.e., have covered a variety of roles including software engineers, developers, consultants, and defense contractors. The other 31.67\% have a less significant professional experience in the field and, after an undergraduate degree in a computing-related discipline, have worked either in close fields or have just research or teaching assistant experience.
The remaining 20\% of the students are ``career changers'', i.e., students who have an undergraduate degree in a non-computing related discipline. They usually have some working experience in their field, and have transitioned to computing through a certificate in which they have learned programming, algorithms, computing foundations, and software engineering. The cohort of participants thus includes both professionals and novices.

\subsection{Research Questions (RQs)}
The following RQs are addressed in our study. 
\begin{itemize}
    \item \textbf{RQ1:} \textit{What are the strategies adopted for the selection of inspirational apps to support ASE?} 
    \item \textbf{RQ2:} \textit{What are the benefits and challenges of ASE?}
    \item \textbf{RQ3:} \textit{What are the differences between ASE and IBE?}
\end{itemize}

\begin{figure}[h]
\includegraphics[width=\columnwidth]{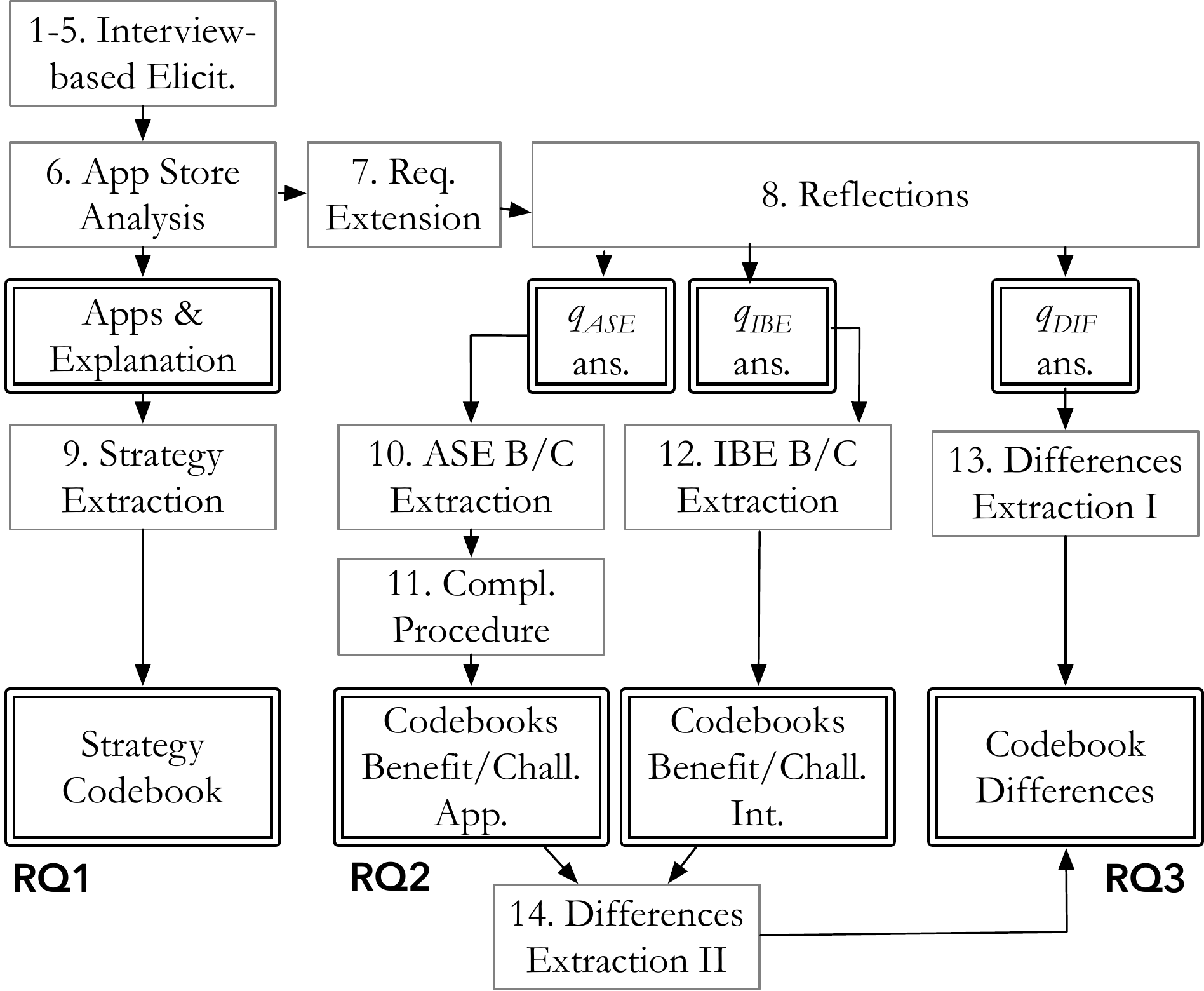}
\caption{Data Collection and Analysis Procedures.}
\label{fig:design}
\end{figure}

The steps of the data analysis and collection procedure (approved by the Institutional Review Board of \university{}) are described in the following, and depicted in Fig.~\ref{fig:design}. Tasks are depicted with single margin, while double margin indicates data. Tasks 1---8 concern activities for data collection, while 9---14 are for data analysis.

\subsection{Data Collection}

Tasks 1---5 concern IBE, including all the activities that go from initial customer ideas to documented requirements. These steps are based on the design of the experiment on elicitation interviews by Debnath \textit{et al.}~\cite{debnath21}. Tasks 6---8 concern ASE. 

\textbf{1. Preparation} Analysts are given a brief description of an app to develop and are asked to prepare questions for a customer that they will 
interview to elicit the products’ requirements. The product is an app for the management of summer camps.
A fictional customer is required to study a set of about 50 user stories, 
which are regarded as the initial customer ideas for the experiment. The user stories are taken from the dataset by Dalpiaz~\cite{dalpiaz2018}, file \texttt{g21.badcamp.txt}. The use of the same customer for all the interviews is in line with similar experiments, such as \cite{bano2019teaching} and \cite{Pitts2004Stopping}. 

\textbf{2. Interview} 
Each analyst performs a 15 minutes interview with the customer, possibly asking additional questions with respect to the ones that they prepared. The  customer answers based on the set of user stories that describe the product---which are not shown to the analysts, to increase realism. 
The analysts are required to record their interviews, and take notes. 

\textbf{3. Requirements Analysis}
Based on the recording and their notes, the analysts have to: (a) perform an initial analysis of the requirements, and based on this analysis (b) produce additional questions for the customer to be asked in a follow-up interview. 

\textbf{4. Follow-up Interview}
The analysts perform a follow-up interview with the customer, which also lasts 15 minutes, to ask the additional questions prepared. 

\textbf{5. Requirements Documentation}
After the second interview, they are required to write down from 50 to 60 user stories for the system. We constrain the number of user stories between 50 and 60 to be consistent with the number of user stories in the original set. About 50 is also the typical number in the dataset by Dalpiaz~\cite{dalpiaz2018}, which we deem representative of user story sets used for research purposes.

\textbf{6. Analysis of App Stores}
To explore and get inspired by possible competing products available on the market, the analysts are asked to perform an informal market analysis based on the app stores, comprising the following steps:

\begin{itemize}
    \item Select at least 5 mobile apps from Google Play or the Apple App Store that are in some way related to the developed product (e.g., other apps for summer camps, apps for trekking, or anything that they consider related).
    \item  Try out the selected apps to have an idea of their features when compared with their product. 
    \item Go through the app reviews to identify desired features, and additional requirements that may be appropriate also for their product. 
\end{itemize}

No automated tool, except for the default app store search engines, was provided for the market analysis task, and the analysts were free to browse the app stores following their intuition. The goal was to avoid confounding factors, i.e., the usage of a tool, and also to elicit challenges that could be addressed through tools, including possibly existing ones. 
Based on the analysis of the app stores, the analysts are required to list the selected apps and their links, together with a brief description that 1. outlines the main features of the product, 2. explains in which way the product is related to the original one, and why they have chosen it.

\textbf{7. Requirements Extension}
Based on the analysis of the app stores, the analysts are asked to add 20 user stories to their original list.

\textbf{8. Reflections}
The analysts are asked to fill out a questionnaire to reflect on the experience. This includes three open-ended questions: $q_{ASE}$) What are the benefits and challenges of ASE?; 
$q_{IBE}$) What are the benefits and challenges of IBE?; 
$q_{DIF}$) What are the differences between IBE and ASE?

\subsection{Data Analysis}
\label{sec:dataanalysis}
Data analysis is performed by two researchers, by means of thematic analysis according to Braun and Clarke~\cite{braun2006using}, within a \textit{critical realist} framework, using a \textit{theoretical} approach (i.e., RQ-driven), and following the guidelines by Salda{\~na} for the coding activity~\cite{saldana2021coding}. 
In the following, we report the analysis performed for each RQ, with examples of how the themes were produced. For RQ2, we also complemented the analysis with a literature review, 
and for RQ3 we included a brainstorming activity to analyse the differences between elicitation styles. For the complete and more detailed protocol, the codebooks, and the coded data, we refer to the supplementary material~\cite{ferrari2022supplementary}.

\textbf{9. RQ1: Strategy Extraction} The input data are the brief descriptions of apps with motivations for the choice, cf. step 7. 
The data from the 58 analysts include 295 items, one for each app. These were coded in multiple iterations of open and closed coding by the researchers. Codes were cross-checked to come to a consolidated set of themes, which were then re-applied to the data. For example, a statement such as ``\textit{This app is related to my app because it is a resource for people who are interested in camping and other outdoor events and activities}'' was associated with the theme \textit{user profile similarity}. At the end of the activity, the researchers redacted a codebook including: strategy name; description; example cases.  


\textbf{10. RQ2: ASE Benefits/Challenges Extraction} The input data are the reflections from step 8, answers to $q_{ASE}$. An iterative process of open and closed coding was followed also in this case. For example, the item containing the text ``\textit{getting hands-on experience with potential features and evaluating their implementation}'' was coded among benefits as \textit{hands-on evaluation}. Instead, ``\textit{User reviews are a majority either based on total hatred of a product or complete satisfaction. These two extremes do not lead to productive identification of what requirements are missing in a product}'' was coded among challenges as \textit{polarised reviews}. The researchers aggregated the codes into common categories (i.e., higher-level themes), and produced two codebooks, one for benefits, and one for challenges, with: category; benefit/challenge; description; examples. 


\textbf{11. RQ2: Complementary Procedure} to complement benefits and challenges of ASE, we performed a lightweight systematic literature review (SLR) following the guidelines by Kitchenham~\cite{kitchenham2004procedures}.  Specifically, we queried the Scopus engine with the following string (( benefit*  OR  challenge*)  AND  ( (app  AND  store*)  OR  (app  AND  review*)) on title, abstract, and keywords, selecting only the Computer Science subject area, and specific venues (e.g., IEEE TSE, Springer EMSE). The full string is available in our supplementary material~\cite{ferrari2022supplementary}. 
Scopus’ coverage is considered optimal when compared to other databases (e.g., IEEE Xplore, ACM Digital library) ~\cite{martinez2022software}. 
The highly selective search string identified 82 publications. We set as main inclusion criterion ``\textit{The paper mentions one or more benefits or challenges of app store analysis}''. After screening the papers based on this criterion, we finally selected 35 items. These were analysed by the researchers with open coding, to extract further benefits and challenges that could apply to ASE. 

\textbf{12. RQ3: IBE Benefits/Challenges Extraction} The input data are the reflections from step 8, answers to $q_{IBE}$. An analysis analogous to the one described in step 10 was carried out on the answers to $q_{IBE}$, to produce two codebooks with benefits and challenges of IBE. These are used in step 14. We do not perform an SLR in this case, as ASE is our main focus. 

\textbf{13. RQ3: Differences Extraction I} 
Thematic analysis was performed on the reflections from step 8, answers to $q_{DIF}$, to produce a first comparative table, contrasting the characteristics of ASE and IBE.
For example, the sentence ``\textit{interviews are designed to elicit what the stakeholder wants, and product-inspired elicitation is designed to elicit what the stakeholder potentially didn't know he wanted or needed}'' led to two contrasting items: (ASE) ``\textit{identifies what the stakeholder could need}'' vs (IBE) ``\textit{identifies what the stakeholder wants}''. This led 
to a first version of the comparative table.  

\textbf{14. RQ3: Differences Extraction II} To complete the table, the researchers jointly analysed the codebooks for ASE and IBE (four codebooks). 
For each item in each codebook of an elicitation style, they brainstormed on possible differences with the other style, considering the other themes in the other codebooks. For example, in relation to the challenge for IBE previously coded as \textit{conflicting requirements}, the researchers identified the  contrasting theme \textit{conflicting reviews} (cf. Table~\ref{tab:benefitschallenges}). The contrast was reformulated as (IBE) ``\textit{Need to deal with stakeholders' inconsistencies}'' vs (ASE) ``\textit{Need to deal with user review inconsistencies}''. 
The activity led to the final version of the comparative table (Table~\ref{tab:tablecomparison}). 

\section{Results}
\label{sec:results}

\subsection{RQ1: Strategies}

In the following, we report the set of different strategies identified, together with representative quotes from our data (in \textit{italic}). The strategy for the selection of an app is not unique, in the sense that multiple strategies could be combined to drive the selection of a certain app. To better understand the quotes, it is useful to report the features of the original app idea. This is an app for the management of summer camps, including the following features: (1) managing the information, registration, and activities of the participants, (2) giving the participants' guardians the opportunity to register and follow their children, (3) managing the employees’ performance and schedule, (4) communicating with parents and employees, and (5) social multi-media features. 
The following strategies are identified.
\begin{itemize}

\item \textbf{similarity by functionality enhancement}: the app implements \textit{only one} functionality that is considered to be missing, or not sufficiently developed,  in the tool. To select these apps, one can assume that analysts strategically considered the elicited requirements, and searched for products implementing specific functionalities, e.g., 
\textit{``Sling is an app to manage employee scheduling. [...] 
This app is related to my product through its messaging and schedule building features.''} 

\item \textbf{lexical similarity}: the app is considered because it has been likely returned by the available search engines with straightforward app-related keywords (``summer camp'' or ``camping'', in our case) but does not have much to do with the original app. This rationale is not explicitly stated by the analysts, but it is visible in the following statement, 
in which the identified similarity is rather minimal: 
     ``\textit{The Dyrt is a camping app that lets you get access to camping information in the US. [...] This app is related to the developed product because there is a feature for you to create a profile and also leave reviews about the campground}''. The Dyrt is one of the first apps retrieved when querying Google Play with ``camping''.

\item \textbf{similarity by functionality subset}: only one specific subset of functionalities of the app is considered as a possible inspiration. 
This case frequently occurred when the analysts retrieved an app based on lexical similarity and then found that certain functionalities could be considered. For example, considering the app \textit{Recreation.gov} (again a camping app) an analyst said: \textit{``Regarding the system being created, allowing the users to 
select what activities they want would be beneficial. 
They can see photos of the activities and what they will be 
doing [...]. The registration will be secure when the campers make their account, and they will 
be able to see real-time feed''}.


\item \textbf{domain similarity}: the app is considered because it belongs to a similar application domain, e.g.,  hiking, orienteering, outdoor games. The domain similarity of a certain app can enable the identification of interesting functionalities not initially planned. For example, \textit{Cairn}, a hiking app to keep users safe and connected, allows users to ``\textit{set the paths that they are taking and share them with friends and family. If they are not back from the hike by a certain time that the user has chosen, the app will send a notification to their friends}''.    

\item \textbf{common use} the app is well-known and commonly used by a large user base (e.g. Facebook, Instagram), and it has been likely chosen without searching the app store. For example, \textit{Garmin} is considered because 
``\textit{[the stakeholder] wants a means to track guests and staff while on the campgrounds. The Garmin wearable allows for the device to transmit GPS coordinates which can be used for tracking}''.

\item \textbf{user profile similarity}: as exemplified in Section~\ref{sec:dataanalysis} (step 9), the app is selected because the profile or interest of potential users is considered similar. 
    
    

\item \textbf{similarity by software scope}: the app is domain-agnostic or belongs to a completely different domain, but it has a similar general scope (e.g., management). Considering these apps can help to build a generic product, but also to enhance existing features: 
    ``\textit{This app allows users to create custom business apps for yourself and your team. The app comes with templates for inventories, invoicing/accounting, [...] 
    it has so many of the features necessary to manage a business}''. 

     \item \textbf{generic product}:	the app belongs to the same or similar application domain, and it is a context-independent product of the specific software, specifically designed to reach a broader audience. These apps can be particularly useful to identify generic requirements, which cannot be gathered through interviews in single specific contexts: 
     ``\textit{The application is customized for each camp site and requires specific log on information. This is an additional requirement for this application to be marketable to many other businesses, and certainly would not have been covered as part of the interviews as the interviewee has no interest in a product for anyone else}''.

    \item \textbf{similarity by business mission}: the app is similar because the overall goal of the \textit{business} is similar, e.g., education. Here, the similarity is not specifically driven by the software features, but by the mission of the actual business supported by the software. For example, in the app of the daycare \textit{Kriyo School}, ``\textit{if you are an administrator of  a daycare or preschool you can create an account to manage any aspect of the daycare. This could be the broader audience for the product.''} 

    

    
     \item \textbf{full match}: the app belongs to the exact same application domain, and it is implementing the same functionalities. Looking at these apps enables analysts to mimic certain features. 
     \textit{CampMinder}, an application specific for summer camp management, is a full match, as ``\textit{it allows for users to access the records associated with the children in their camp and it specifically integrates online registration and forms}''. 

\item \textbf{popularity} the app is selected because it appears to be widely used or highly rated/awarded. For example, as 
``\textit{there is no better tool for managing customers than a good CRM [Customer Resource Manager]}'', some analysts select CRMs. Among them, \textit{HubSpot} has been selected as it ``\textit{is a fairly famous CRM product that I think would have good features to get inspired by}''.

 \end{itemize}
 
It should be noted that, while in some cases the analysts strictly relied on the output of the app search engines on the basis of simple app-related queries (e.g., \textit{functionality subset, lexical, domain, generic product}), in other cases they appear to have made an effort to think and search more strategically based on a rationale for extension of their product (e.g., \textit{functionality enhancement, common use}).  

\begin{table}[]
\centering
\resizebox{0.8\columnwidth}{!}{

\begin{tabular}{ll}
\hline
\multicolumn{1}{c|}{\textbf{Benefits}}                                                                         & \multicolumn{1}{c}{\textbf{Challenges}}                                                             \\ \hline
\multicolumn{2}{c}{\textbf{CONCEPT}}                                                                                                                                                                                 \\ \hline
\multicolumn{1}{l|}{\textit{\textbf{Concept   Inspiration}}}                                                   & \textit{\textbf{Product Concept}}                                                                   \\
\multicolumn{1}{l|}{inspiration   for ideas}                                                                   & limitation of creativity                                                                            \\
\multicolumn{1}{l|}{broaden   vision}                                                                          & loss of original purpose                                                                            \\
\multicolumn{1}{l|}{}                                                                                          & no support for initial idea                                                                         \\
\multicolumn{1}{l|}{}                                                                                          & risk to copy products                                                                               \\ \hline
\multicolumn{2}{c}{\textbf{REQUIREMENTS}}                                                                                                                                                                            \\ \hline
\multicolumn{1}{l|}{\textit{\textbf{Requirements   Inspiration}}}                                              & \textit{\textbf{\begin{tabular}[c]{@{}l@{}}Requirements and Features   \\ Definition\end{tabular}}} \\
\multicolumn{1}{l|}{identify   novel features}                                                                 & difficult to understand goals                                                                       \\
\multicolumn{1}{l|}{expanding   requirements}                                                                  & irrelevant requirements                                                                             \\
\multicolumn{1}{l|}{narrowing   requirements}                                                                  & relevant/key feature selection                                                                      \\
\multicolumn{1}{l|}{identify   unknown requirements}                                                           & difficult adaptation of features                                                                    \\
\multicolumn{1}{l|}{\begin{tabular}[c]{@{}l@{}}overcoming   the lack of \\      domain knowledge\end{tabular}} & \begin{tabular}[c]{@{}l@{}}risk to miss relevant  \\  requirements\end{tabular}                     \\
\multicolumn{1}{l|}{}                                                                                          & \begin{tabular}[c]{@{}l@{}}identification of demographic   \\ preferences*\end{tabular}             \\ \hline
\multicolumn{2}{c}{\textbf{PRODUCT}}                                                                                                                                                                                 \\ \hline
\multicolumn{1}{l|}{\textit{\textbf{Implementation   Inspiration}}}                                            & \textit{\textbf{\begin{tabular}[c]{@{}l@{}}Product Search and \\ Evaluation\end{tabular}}}          \\
\multicolumn{1}{l|}{issue   identification}                                                                    & \begin{tabular}[c]{@{}l@{}}difficult to identify relevant \\ products\end{tabular}                  \\
\multicolumn{1}{l|}{leverage   developers knowledge}                                                           & too many similar products                                                                           \\
\multicolumn{1}{l|}{hands-on   evaluation}                                                                     & insufficiency of the app store                                                                      \\
\multicolumn{1}{l|}{inspiration   for implementation}                                                          & static categorisation*                                                                              \\
\multicolumn{1}{l|}{enhancing   product}                                                                       & \begin{tabular}[c]{@{}l@{}}need to log-in to evaluate   \\ products\end{tabular}                    \\
\multicolumn{1}{l|}{create   adaptable product}                                                                & need to purchase app functions                                                                      \\
\multicolumn{1}{l|}{identify   irrelevant elements}                                                            & problems with ads*                                                                                  \\
\multicolumn{1}{l|}{}                                                                                          & repackaged apps*                                                                                    \\ \hline
\multicolumn{2}{c}{\textbf{USER}}                                                                                                                                                                                    \\ \hline
\multicolumn{1}{l|}{\textit{\textbf{Market   Satisfaction}}}                                                   & \textit{\textbf{Reviews (Quality)}}                                                                 \\
\multicolumn{1}{l|}{identify   market needs}                                                                   & uninformative reviews                                                                                        \\
\multicolumn{1}{l|}{identify   market trends}                                                                  & vague/poor reviews                                                                                       \\
\multicolumn{1}{l|}{\begin{tabular}[c]{@{}l@{}}create   competitive \\      product\end{tabular}}              & short reviews                                                                                       \\
\multicolumn{1}{l|}{identify   successful solutions}                                                           & unstructured reviews*                                                                               \\ \cline{1-1}
\multicolumn{1}{l|}{\textit{\textbf{User   Satisfaction}}}                                                     & fake reviews*                                                                                       \\ \cline{2-2} 
\multicolumn{1}{l|}{improve   usability}                                                                       & \textit{\textbf{Reviews (Content)}}                                                                 \\
\multicolumn{1}{l|}{improve   accessibility*}                                                                  & polarised reviews                                                                                   \\
\multicolumn{1}{l|}{user   feedback}                                                                           & limited constructive criticism                                                                      \\
\multicolumn{1}{l|}{assess   usability}                                                                        & conflicting reviews                                                                                 \\
\multicolumn{1}{l|}{consider   large audience}                                                                 & partial information*                                                                           \\
\multicolumn{1}{l|}{satisfy   variety of users}                                                                & \begin{tabular}[c]{@{}l@{}}reviews do not comment \\ features\end{tabular}                          \\ \cline{2-2} 
\multicolumn{1}{l|}{identify   user values}                                                                    & \textit{\textbf{Reviews (Quantity)}}                                                                \\
\multicolumn{1}{l|}{}                                                                                          & uncategorised reviews*                                                                              \\
\multicolumn{1}{l|}{}                                                                                          & too many reviews                                                                                    \\
\multicolumn{1}{l|}{}                                                                                          & insufficient number of reviews                                                                      \\
\multicolumn{1}{l|}{}                                                                                          & filtering reviews                                                                                   \\ \hline
\multicolumn{2}{c}{\textbf{PROCESS}}                                                                                                                                                                                 \\ \hline
\multicolumn{1}{l|}{\textit{\textbf{Process   Support}}}                                                       & \textit{\textbf{Process Weaknesses}}                                                                \\
\multicolumn{1}{l|}{}                                                                          & time consuming activity                                                                             \\
\multicolumn{1}{l|}{limited   stress}                                                                          & unfocused activity                                                                                  \\
\multicolumn{1}{l|}{save   development time}                                                                   & difficult to trace stakeholders                                                                     \\
\multicolumn{1}{l|}{idea   validation}                                                                         & requirements validation                                                                             \\
\multicolumn{1}{l|}{assess   feasibility}                                                                      & intellectual property issues                                                                        \\
\multicolumn{1}{l|}{support   decision process}                                                                & no follow-up                                                                                        
\end{tabular}
}
\caption{Benefits and Challenges of ASE}
\label{tab:benefitschallenges}
\end{table}

\subsection{RQ2: Benefits and Challenges}

Table~\ref{tab:benefitschallenges} reports the summary of the results for RQ2. The superscript * indicates items that were identified through the SLR. In the following, we consider the main categories, and present representative themes, with associated example quotes. 

\subsubsection{Benefits} these are divided into \textit{concept inspiration}, \textit{requirements inspiration}, \textit{implementation inspiration}, \textit{user satisfaction}, \textit{market satisfaction} and \textit{process support}.

\paragraph{Concept Inspiration} looking at a variety of different products can generate novel ideas for adaptations of the app and broaden the scope and  \textit{vision} of the initial idea~\cite{karras2020representing}, leading to a transformation of the original product to satisfy a possibly different market. In this regard, one of the analysts said ``\textit{[ASE] gave me a different outlook on how to gather new ideas}'' (\textbf{inspiration for ideas}).

\paragraph{Requirements Inspiration} looking into other products helps to identify new possible features, extend the existing requirements, or better define lower-level ones, thanks to the possibility of analysing product implementations:  ``\textit{best features that are successful in other products can be brought in, new ideas can be implemented by looking at other products}'' (\textbf{identify novel features}).

Looking at the market can also help analysts to overcome their lack of domain knowledge, which is a central issue in requirements elicitation activities~\cite{Hadar2014Role,NiknafsB17}, and can possibly lead to the identification of tacit/unknown requirements~\cite{gervasi2013unpacking,ferrari2016ambiguity}. 
For example, one analyst said that ASE ``\textit{can help to shape and provide more specific requirements. It is focused on the application functionality and could also help when there is not a lot of domain knowledge about the product}'' (\textbf{overcoming the lack of domain knowledge}).

\paragraph{Implementation Inspiration} ASE can provide ideas to guide the development phase. Problems with certain implementations can be easily identified, by looking at other products and their users' feedback, which ``\textit{can show you issues users had with these products, so you can attempt to avoid those issues when designing your product}'' (\textbf{issue identification}). Identification of common issues is particularly useful, as many apps tend to use similar libraries~\cite{mojica2013large}. Irrelevant features and enhancements of existing ones can be discovered, as well as possible implementation solutions. To this end, \textit{hands-on} evaluation of existing products is a most valuable support. Indeed, 
\textit{``You can even try out the functionality to see how things look and feel.''} 
(\textbf{hands-on evaluation}).
ASE can also help to implement a product that is more adaptable to different needs, as products in the app store are generally oriented towards a broad audience:``\textit{[ASE] gives us a bigger picture of how the features could be implemented and what other requirements could be developed later so the system would be ready for such features later.''} (\textbf{create adaptable product}). 
Overall, analysts and developers can build upon the knowledge of other developers through their implemented solutions, thereby overcoming their possible limitations in terms of competences. 








\paragraph{User Satisfaction} trying out other products and checking their reviews can help to assess whether certain solutions are usable or not, possibly preventing usability issues: \textit{``as I previewed some of the software currently available, I can see how the interfaces may be easier to use''} (\textbf{assess usability}). 
Usability comments are frequently associated with lower ratings with respect to other types of reviews~\cite{chen2021should}. 
Also, looking at user feedback can also identify accessibility issues~\cite{reyes2022accessibility}. In addition, user feedback informs the analyst on what is required and what is not needed by users of similar apps, and even infer what are user \textit{values}, which are related to the core benefits that one expects from existing apps:  
``\textit{the end-goal should not be about how well they created something; it should be about how much value this brings to the customer}'' (\textbf{identify user values}). 
As suggested by Sutcliffe \textit{et al.}~\cite{sutcliffe2021investigating}, user values can motivate the download and use of certain apps. 


Finally, looking at similar products also helps to understand what are the different types of potential user profiles, and how to satisfy them. 
With ASE, ``\textit{it becomes possible to create a system that is more likely to meet the needs of a variety of users}'' (\textbf{satisfy variety of users}). User profiling is recognised as particularly important for personalised applications, especially when they include some form of content recommender system~\cite{zanker2019measuring}, e.g., music or video apps.






\paragraph{Market Satisfaction} by looking into apps and their reviews, analysts can better identify what is required by the market: \textit{``[ASE] gives an opportunity to understand what current systems are fulfilling the user needs and also the unfulfilled users requirements.''}
(\textbf{identify market needs}).  
By monitoring the market trends, 
one can understand how the market evolves, thus anticipating future needs. Looking at other products satisfying similar requirements can be a reality test to check whether the product idea is sufficiently in line with the market expectations, and in which way it needs to be improved to achieve a competitive advantage: \textit{``I feel that if I had the opportunity to implement these features into the product that it would have a fair chance competing in the app marketplace}'' (\textbf{create competitive product}). In this regard, the analysis of successful solutions, based on the number of downloads and positive reviews---typically considered as indicators of popularity~\cite{martin2015app}---, can be particularly helpful to understand what is the current benchmark.

\paragraph{Process Support} different phases and aspects of the development process can be supported by ASE. Specifically, it can help during the initial validation of previous ideas, and also to assess their feasibility. If similar products have implemented certain solutions, these should not pose insurmountable technical barriers:  ``\textit{The benefit of ASE is that you get your requirement off a product that already exists so you know that whatever requirement gotten from this process is feasible for the intended product}'' (\textbf{idea validation}; \textbf{assess feasibility}). 
Since there is no interaction with stakeholders, ASE is not considered stressful, and can also save development time by helping to estimate it. Indeed, ``\textit{building something from scratch is hard and time inefficient, so  by taking a look at what is out there one can gauge factors such as development time''} 
(\textbf{save development time}). 
 From the development standpoint, understanding what is relevant in competing products can facilitate the prioritisation of features and better direct the decision process towards implementation. 







\subsubsection{Challenges} these are divided into \textit{product concept}, 
\textit{requirements and features definition},
\textit{product search and evaluation},
\textit{reviews}, and \textit{process weaknesses}.

\paragraph{Product concept}
while ASE can be useful to extend a product idea, it does not provide sufficient support when one needs to define a novel concept from scratch. 
As noted by one of the analysts, ``\textit{if we are building an application from scratch, I think it is not that useful as we are still working on building the basic functionality of the product. 
}'' (\textbf{no support for initial idea}).
Furthermore, looking at other products can lead to a limitation of creativity, and also to the risk of copying other products:  
``\textit{The danger of ASE is to avoid simply copying another product that exists. You want to design a product that is unique to your customer}''
(\textbf{risk to copy products}).
 Finally, by integrating more and more features adapted from other products, one could lose the original purpose of the app under development. 





\paragraph{Requirements and features definition}

while ASE can drive implementation, it provides little help in understanding possible high-level goals of the stakeholders, as what is available is only the software and its reviews. 
Given the absence of well-defined goals, it is difficult to understand what are the relevant, key features to select, and, without the stakeholders at hand, there is a risk of missing relevant requirements. 
In addition, after selecting features, it is also hard to adapt  and integrate them into an existing product feature set, as 
``\textit{it’s necessary to try the products and understand how the products work. It’s the understanding that helps SE determine how features can be modified and adapted to work for their clients most effectively}''
(\textbf{difficult adaptation of features}). 
Demographic preferences are also hard to identify, especially when one aims to develop an app for different countries~\cite{lim2014investigating}. 






\paragraph{Product search and evaluation}
the app store is not designed to specifically support ASE, and it does not offer a structured way for comparison, as it happens, e.g., on the Amazon marketplace.
A product search typically returns a set of too many similar products, with several features to compare. Furthermore, their categorisation is static, and not in line with the evolution of the market~\cite{ebrahimi2021classifying}. Additional confusion to the search results is introduced by repackaged apps~\cite{li2019rebooting}.
The  information overload makes it difficult to identify which products can be relevant for inspiration. On this topic, one of the analysts said ``\textit{The challenge I faced [during ASE] is that there are not many applications available on the App Store in the camping genre. So, I had to do some intense research, surf many websites, and find links to applications from the websites, which took a lot of time}'' (\textbf{insufficiency of the app store}). 
Evaluating products often requires creating an account to log-in, which complicates the evaluation. Furthermore, the test of many relevant features often requires a subscription to a premium account, thus making the comparison of several products a costly activity from the economic standpoint, or an incomplete one in case of limited resources. When one uses free app versions, fair evaluation is often complicated by an excessive number of ads~\cite{gao2022understanding}.






\paragraph{Reviews} reviews pose challenges in terms of quality, content, and quantity. Specifically, they are often vague and superficial, poorly written, short, and thus not informative. Their format is unstructured, which makes them hard to analyse~\cite{tavakoli2018extracting}. 
One of the analysts, discouraged by the quality of the reviews stated: 
``\textit{one must sift through the countless valueless comments to find the gold nuggets that are missing, nonfunctional, or unnecessary features. People are naturally lazy, and their reviews and comments normally reflect that}'' (\textbf{uninformative reviews}).
In terms of content, reviews are often highly polarised with enthusiastic comments or extremely negative ones. 
The former usually praise the product and do not contain any useful suggestions and the latter often do not include constructive criticism that can be exploited by analysts or developers. 
The polarisation also leads to conflicting reviews, and it is hard to understand which opinion is more trustworthy, also due to the problem of fake reviews~\cite{martens2019towards}. 
On review inconsistency, one of the analysts said ``\textit{ASE has a lot to do with what you personally see and feel plus the advice and comments from the masses. This can be hard to filter through because people may not always know what they want or there are so many conflicting reviews}'' (\textbf{conflicting reviews}).
Also, reviews tend to comment on the product as a whole, and it is hard to find reviews that comment on features, thereby making tools for associating user sentiment to features in-app reviews particularly useful~\cite{guzman2014users}. 
The quantity of reviews is also an issue, especially combined with their quality. Filtering tools that can select only relevant reviews based on a certain query (e.g., Appbot) can provide valuable support, as they help to identify review categories on-demand, and address the problem of uncategorised reviews~\cite{mcilroy2016analyzing}.  
In some cases, the number of reviews can also be insufficient to get an idea of the product. This happens especially if one wants to develop a system for a rather limited market---such as in our experimental simulation---where similar apps may not be sufficiently popular to have a substantial volume of reviews. 
Both extremes in the number of reviews can be a problem: ``\textit{Depending on the product’s popularity, reviews can be insufficient or abundant. When there are a lot of reviews 
you need to find the right ones that are helpful and are not biased and blatant}'' (\textbf{too many reviews}; \textbf{insufficient number of reviews}).
In addition, even when high-quality reviews are available, these always include partial information to make sense of the raised issues, and need to be complemented with external sources such as app crash reports, tweets, community blogs and code repositories~\cite{genc2017systematic,nayebi2018app,palomba2017recommending}.












\paragraph{Process weaknesses} searching, selecting, and analysing products are quick activities compared to developing prototypes, but they unavoidably require time. Since there are no structured guidelines, they also tend to be unfocused, without a linear or directed process. 
For example, an analyst pointed out that ASE ``\textit{is a time-consuming process that requires hours of research to find the best products to compare and then even longer to comb through the publicly available reviews for the given product. }''
(\textbf{time-consuming activity};  \textbf{unfocused activity}).
It is also difficult to understand who are the stakeholders who can use a certain product. 
As stated by one of the analysts ``\textit{the requirement you get [from ASE] cannot be traced to the accurate stakeholder so that you can have a better understanding of why the requirement is needed}'' (\textbf{difficult to trace stakeholders}).
In addition, one cannot have follow-up questions with the users who left an unclear review, or with those that appear to have something more to say about, e.g., a certain bug and its reproduction. 
Since the stakeholders cannot be contacted for in-depth interactions, requirements validation is also not possible. This has been noted by some analysts as one of the biggest challenges with ASE: ``\textit{my biggest challenge in applying this type of elicitation is validating the requirements. I had to make decisions based on what I have received from the user. Therefore, any inclusion of ambiguous requirements will lead to modeling and production of a bad product}'' (\textbf{requirements validation}). Borrowing features from other products can also potentially lead to intellectual property issues, which product developers could raise after the app is released to the public. 







\subsection{RQ3: Differences}

\begin{table*}[]
\centering

\resizebox{0.7\textwidth}{!}{

\begin{tabular}{|llll|}
\hline
\multicolumn{2}{|c|}{\textbf{Interview-based Elicitation}}                                                      & \multicolumn{2}{c|}{\textbf{App Store-inspired Elicitation}}                          \\ \hline
\multicolumn{4}{|c|}{\textbf{Main   focus}}                                                                                                                                                             \\ \hline
\multicolumn{1}{|l|}{+}      & \multicolumn{1}{l|}{Focus   on goals (WHY)}                                      & \multicolumn{1}{l|}{Focus   on solutions (HOW)}                              & +      \\ \hline
\multicolumn{1}{|l|}{+}      & \multicolumn{1}{l|}{Focus   on asking}                                           & \multicolumn{1}{l|}{Focus   on observing}                                    & +      \\ \hline
\multicolumn{1}{|l|}{+}      & \multicolumn{1}{l|}{Focus   on future development (forward thinking)}            & \multicolumn{1}{l|}{Focus   on past development (backward thinking)}         & +      \\ \hline
\multicolumn{1}{|l|}{+}      & \multicolumn{1}{l|}{Personalised   product oriented so satisfy a customer}       & \multicolumn{1}{l|}{General   product oriented to satisfy market needs}      & +      \\ \hline
\multicolumn{1}{|l|}{+}      & \multicolumn{1}{l|}{Support   problem decomposition}                             & \multicolumn{1}{l|}{Support   idea generation}                               & +      \\ \hline
\multicolumn{1}{|l|}{+}      & \multicolumn{1}{l|}{Appropriate   for initial development stage}                 & \multicolumn{1}{l|}{Appropriate   for later development stages}              & +      \\ \hline
\multicolumn{4}{|c|}{\textbf{Elicitation   of needs}}                                                                                                                                                   \\ \hline
\multicolumn{1}{|l|}{+}      & \multicolumn{1}{l|}{Relies   on direct stakeholder interaction}                  & \multicolumn{1}{l|}{Relies   on analysis of user feedback}                   & +      \\ \hline
\multicolumn{1}{|l|}{+}      & \multicolumn{1}{l|}{Explicit   questions about details}                          & \multicolumn{1}{l|}{Collection   of details is incidental}                   & -      \\ \hline
\multicolumn{1}{|l|}{$\sim$} & \multicolumn{1}{l|}{Relies   on proper questions}                                & \multicolumn{1}{l|}{Relies   on proper search queries}                       & $\sim$ \\ \hline
\multicolumn{1}{|l|}{+}      & \multicolumn{1}{l|}{Mainly   qualitative information}                            & \multicolumn{1}{l|}{Qualitative   and quantitative information from reviews} & +      \\ \hline
\multicolumn{1}{|l|}{+}      & \multicolumn{1}{l|}{Can   help to collect stakeholders' goals}                   & \multicolumn{1}{l|}{Stakeholders'   goals are hard to identify}              & -      \\ \hline
\multicolumn{1}{|l|}{+}      & \multicolumn{1}{l|}{Explicit   questions to stakeholders}                        & \multicolumn{1}{l|}{Questions   to stakeholders not possible}              & -      \\ \hline
\multicolumn{4}{|c|}{\textbf{Interpretation   of needs}}                                                                                                                                                \\ \hline
\multicolumn{1}{|l|}{+}      & \multicolumn{1}{l|}{Identify   what the stakeholder wants}                       & \multicolumn{1}{l|}{Identify   what the stakeholders could need}             & +      \\ \hline
\multicolumn{1}{|l|}{+}      & \multicolumn{1}{l|}{Requirements   within the project scope}                     & \multicolumn{1}{l|}{Can   lead to out of scope requirements}                 & -      \\ \hline
\multicolumn{1}{|l|}{+}      & \multicolumn{1}{l|}{Probing/follow-up   questions can reduce misinterpretations} & \multicolumn{1}{l|}{Reviews   can be misinterpreted}                         & -      \\ \hline
\multicolumn{1}{|l|}{+}      & \multicolumn{1}{l|}{Explicit   answers can remove wrong assumptions}             & \multicolumn{1}{l|}{Wrong   assumptions can be made about user needs}        & -      \\ \hline
\multicolumn{1}{|l|}{+}      & \multicolumn{1}{l|}{Understand   product goals and vision}                       & \multicolumn{1}{l|}{Create   product goals and vision}                       & +      \\ \hline
\multicolumn{1}{|l|}{$\sim$} & \multicolumn{1}{l|}{Need   to deal with stakeholders' inconsistencies}           & \multicolumn{1}{l|}{Need   to deal with user review inconsistencies}         & $\sim$ \\ \hline
\multicolumn{1}{|l|}{+}      & \multicolumn{1}{l|}{Success   depends on agreed criteria}                        & \multicolumn{1}{l|}{Success   also depends on luck factors}                  & -      \\ \hline
\multicolumn{4}{|c|}{\textbf{Relationship   with stakeholders}}                                                                                                                                         \\ \hline
\multicolumn{1}{|l|}{+}      & \multicolumn{1}{l|}{Fosters   stakeholders' relationship}                        & \multicolumn{1}{l|}{Limited   relationship with stakeholders}                & -      \\ \hline
\multicolumn{1}{|l|}{+}      & \multicolumn{1}{l|}{Can   change stakeholders' viewpoint}                        & \multicolumn{1}{l|}{Does   not affect stakeholders' viewpoint}               & -      \\ \hline
\multicolumn{1}{|l|}{+}      & \multicolumn{1}{l|}{Can   exploit non-verbal cues}                               & \multicolumn{1}{l|}{No   face-to-face interaction}                           & -      \\ \hline
\multicolumn{1}{|l|}{+}      & \multicolumn{1}{l|}{Allow   access to multiple stakeholders}                     & \multicolumn{1}{l|}{Access   only to users}                                  & -      \\ \hline
\multicolumn{1}{|l|}{}       & \multicolumn{2}{c|}{\textbf{Process}}                                                                                                                           &        \\ \hline
\multicolumn{1}{|l|}{+}      & \multicolumn{1}{l|}{Structured   process}                                        & \multicolumn{1}{l|}{Unstructured   process}                                  & -      \\ \hline
\multicolumn{1}{|l|}{$\sim$} & \multicolumn{1}{l|}{Requires   soft skills}                                      & \multicolumn{1}{l|}{Requires   technical skills}                             & $\sim$ \\ \hline
\multicolumn{1}{|l|}{-}      & \multicolumn{1}{l|}{Stakeholder-directed   process}                              & \multicolumn{1}{l|}{Creative   process}                                      & +      \\ \hline
\multicolumn{1}{|l|}{-}      & \multicolumn{1}{l|}{Stressful   activity}                                        & \multicolumn{1}{l|}{Not   stressful activity}                                & +      \\ \hline
\multicolumn{1}{|l|}{-}      & \multicolumn{1}{l|}{Requires   preparation}                                      & \multicolumn{1}{l|}{No   preparation needed}                                 & +      \\ \hline
\multicolumn{1}{|l|}{-}      & \multicolumn{1}{l|}{Requires   time management}                                  & \multicolumn{1}{l|}{No   time constraints}                                   & +      \\ \hline
\multicolumn{1}{|l|}{-}      & \multicolumn{1}{l|}{More   time consuming}                                       & \multicolumn{1}{l|}{Less   time consuming}                                   & +      \\ \hline
\multicolumn{1}{|l|}{-}      & \multicolumn{1}{l|}{Requires   active control of the conversation}               & \multicolumn{1}{l|}{Inherent   control of the analysis process}              & +      \\ \hline
\multicolumn{4}{|c|}{\textbf{Requirements   assessment}}                                                                                                                                                \\ \hline
\multicolumn{1}{|l|}{+}      & \multicolumn{1}{l|}{Requirements   can be validated by the customer}             & \multicolumn{1}{l|}{Difficult   to validate requirements}                    & -      \\ \hline
\multicolumn{1}{|l|}{-}      & \multicolumn{1}{l|}{Evaluation   possible only with prototypes/mockups}           & \multicolumn{1}{l|}{Hands-on   evaluation of requirements implementation}    & +      \\ \hline
\multicolumn{1}{|l|}{-}      & \multicolumn{1}{l|}{Usability   requirements not testable}                       & \multicolumn{1}{l|}{Can   enable the test of usability requirements}         & +      \\ \hline
\multicolumn{1}{|l|}{-}      & \multicolumn{1}{l|}{Feasibility   cannot be assessed}                            & \multicolumn{1}{l|}{Can   enable assessment of feasibility}                  & +      \\ \hline
\end{tabular}
}
\caption{Differences between interview-based and app store-inspired elicitation.}
\label{tab:tablecomparison}

\end{table*}

Table~\ref{tab:tablecomparison} reports the identified differences between IBE and ASE, divided into six categories. The table also reports whether a certain characteristic is positive (+), negative (-) or neutral ($\sim$)---this evaluation is arguably made by the authors. 

\paragraph{Main focus} the main focus of IBE is eliciting goals, thus answering \textit{why} questions, while ASE aims to understand what could be the possible solutions to address certain goals and answer \textit{how} questions. To this end, IBE is based on asking explicit inquiries that can inform future development (\textit{forward thinking}), while ABE is based on observing implementations, and thus relies on previously developed apps (\textit{backward thinking}). As pointed out by one of the analysts, ``\textit{Interviews help us during the initial phase of software development [...]
where product-inspired elicitation is useful when we are working on improving the system based on customer feedback, review and bug reporting}'' (\textbf{appropriate for initial development stage} vs \textbf{appropriate for later development stage}). So, interviews are useful to create personalised products and can help the analysts in the initial phases of the development, when they need to perform problem decomposition. Instead, ABE helps to develop novel ideas to create a general product oriented to satisfy market needs.


\paragraph{Elicitation of needs}
concerning this aspect, greater advantages are observed for IBE, compared to ASE. In IBE, needs are directly elicited from stakeholders, and the analyst can ask detailed questions, which need to be well formulated to acquire relevant, and mainly qualitative, information. ASE relies on the analysis of user feedback to understand the customers' needs, and collection of details is incidental and not driven by the analyst's investigation. The technique relies on proper search queries, and can be useful to collect both qualitative and quantitative information (number of downloads, rating). While interviews can help to collect goals, these are harder to identify with ASE, because the analyst cannot pose explicit questions to the stakeholders: \textit{``it [ASE] can inspire new ideas [...], but you will not be able to elicit new goals from product-inspired elicitation''} (\textbf{can help to collect stakeholders' goals} vs \textbf{stakeholders' goals hard to identify}).

\paragraph{
Interpretation of needs} this aspect is facilitated by IBE, while some challenges exist for ASE. IBE helps to identify what the stakeholders need, while ASE helps to interpret what they \textit{could} need. In this sense, the former helps to \textit{understand}, while the latter aims to \textit{create} product goals and vision. With IBE, elicited requirements remain within the project scope thanks to the continuous exchange of information with the stakeholders, where probing/follow-up questions can reduce misinterpretations, and remove wrong assumptions. ASE can lead to requirements that are out of the project scope, and misinterpretations of the users' voice expressed through reviews is likely, possibly leading to wrong assumptions: ``\textit{The other main difference is that it is much easy to gain clarity from the client in an interview [...] In product-inspired elicitation you are making decisions or assumptions based on your interpretation of the product so you may have one understanding of the project and its needs which may not necessarily line up with the client}'' (\textbf{explicit answers can remove wrong assumptions} vs \textbf{wrong assumptions can be made about user needs}). While with IBE application success depends on agreed criteria, with ASE it depends also on luck factors, as the market can be moody, and search engines cannot be fully controlled. Inconsistency is a common pain point between IBE and ASE, with conflicting stakeholder  requirements and conflicting reviews.   

\paragraph{Relationship with stakeholders}
IBE revolves around the creation of rapport and fostering good relationships with the stakeholders, while with ASE the stakeholders are not reachable and no actual relationship can be established: \textit{``The main difference between interviews and product-inspired elicitation, is the fact that in the interview, I was able to get a better connection to the human needing something''} (\textbf{foster stakeholders' relationship} vs \textbf{limited relationship with stakeholders}).   
With IBE one can engage in dialogue and possibly change the stakeholders' viewpoint in case of misunderstanding, which is not possible with ASE.
During dialogues, non-verbal cues, such as facial expressions and body language in general, can be exploited to better reveal the stakeholders' inner feelings and thoughts, while no face-to-face, synchronous interaction is possible with ASE. Finally, reviews used in ASE are typically written by users, while interviews can reach a larger set of stakeholder types, e.g., domain experts, sponsors.   

\paragraph{Process} while IBE can follow a structured, mainly sequential, process based on interview scripts prepared beforehand, with ASE the process is unstructured and iterative, as no guideline exists to perform it. The skills required for the two approaches are different, as IBE requires soft skills to understand and create rapport with stakeholders, while with ASE technical skills are central to understand what certain solutions entail from the development standpoint, and how  they can be incorporated into an existing app: 
``\textit{The main difference is that one is a soft skill, and the other is technical. When interviewing, you have to think on your feet, adjust conversation based on stakeholder needs, adjust, build rapport, and make the stakeholder feel comfortable enough to share real goals with you. Product-inspired elicitation is a technical skill}'' (\textbf{requires soft skills} vs \textbf{requires technical skills}). Concerning other process-related aspects, ASE has several advantages over IBE. Specifically, the IBE process is mainly directed by the stakeholders, and in particular, the customer and the sponsor, while ASE is a creative process. IBE can be also stressful, as interaction with other, unknown people, can create some tension. 
Interviews require preparation, while with ASE one is free to improvise, and no specific planning is necessarily needed. Another crucial difference concerns \textit{time}. The involvement of stakeholders in IBE requires to schedule appointments in advance, and manage time well during the interview. 
Instead, with ASE
the analyst has full control of time and of the analysis process and does not need to depend on others' schedules or actively control the conversation to properly manage time and information acquisition.  

\paragraph{Requirements assessment} this aspect is generally easier with ASE, except for requirements validation. In IBE requirements can be explicitly validated by the customer, while validation of ASE requirements implicitly comes after the product has been released on the market. On the other hand, ASE offers the possibility of hands-on evaluation of requirements implementations, while in IBE one can evaluate requirements only with prototypes/mockups. Furthermore, tests of usability requirements, and assessment of implementation feasibility, can be hard with IBE.   
\section{Discussion}
\label{sec:discussion}


\paragraph{Implication for researchers}

Our results are based on a manual version of ASE (i.e., not supported by tools), and confirm the need for many of the technical solutions that have been developed by researchers. These include app review classifiers~\cite{maalej2016automatic,jha2018using}, fake review detectors~\cite{martens2019towards}, review rationale identifiers~\cite{kurtanovic2018user}, but also the more recent works on app comparison and feature recommendation~\cite{jiang2019recommending,liu2022mining,dalpiaz2019re}, which would be the core of ASE. Together with other works~\cite{palomba2017recommending,al2019app},  this shows that there is a \textit{practical need} for tool support in app store analysis, thus addressing the demand for more evidence in this regard observed by Dabrowski \textit{et al.}~\cite{dkabrowski2022analysing}. While research tools address only one problem at a time, \textit{integrated} platforms that collect multiple capabilities, considering our list of observed challenges, are required to fully exploit the potential of ASE. Furthermore, our analysts observed that this practice can be unfocused, as they were not able to devise an intuitive and structured process to perform it. We are not aware of guidelines for ASE, and researchers are called to define them. In this regard, the identified \textit{strategies} can be taken as a starting point to provide suggestions on how to perform ASE in a fruitful manner, depending on the development stage of the product, e.g., using \textit{user profile similarity} during market positioning or substantial app renewal, or  \textit{functionality enhancement}, when performing minor releases. The guidelines should be integrated into agile processes, as these are the most common in app development~\cite{francese2017mobile,jabangwe2018software}. Still on strategies, experimental evaluations can be carried out to assess which ones are more effective given a certain development goal, e.g., extending a product, or  generalising it for a wider market. Implementing recommender systems in which users can tune the search strategy based on their needs is another research direction. Finally, our work calls researchers to better study traditional elicitation techniques, such as interviews, in combination with ASE. While these practices have been largely studied independently, our results show that they can provide \textit{complementary} contributions to the development.





\paragraph{Implication for practitioners}


The main message concerns the list of \textit{benefits} of practicing ASE, 
which should encourage companies to invest more on it as a part of their development process, and not as a mere complementary activity. Among benefits, developers should primarily consider the assessment of feasibility---to be performed when planning for a certain feature---and usability---to be performed during GUI design, a core aspect of app development~\cite{francese2017mobile}. This suggests that different \textit{stages} of development may profit from ASE. Also, different \textit{roles} may exploit it, as, e.g., requirements analysts (concept inspiration),  advanced developers (implementation inspiration), and novices. Indeed, one of the benefits of ASE is also the possibility of overcoming the lack of knowledge, by leveraging the competence of other developers. The activity can thus be particularly useful for young developers and can be potentially exploited as an onboarding exercise for companies, given the need for appropriate strategies in this regard, as observed by Britto \textit{et al.}~\cite{britto2020evaluating}. Training should push for the adoption of planned strategies for app search and selection, rather than clerical usage of search engines, as some of our analysts chose to do. 
App developers need also to be aware of ASE limitations, which can be addressed through IBE. For example, the possibility of explicitly eliciting goals and removing wrong assumptions. IBE can be applied both at the initial development stage, and later, to confirm the unclear feedback coming from reviews. ASE can then be fully exploited to generalise the app idea to more users. In case both strategies are used, practitioners should consider that different analyst profiles can be more appropriate, having soft-skills for interviewers, and technical skills for ASE analysts. 
\section{Threats to Validity}
\label{sec:threats}

\paragraph{Construct Validity} we used thematic analysis to analyse the data according to Braun and Clarke~\cite{braun2006using}, and adopted the guidelines of {Salda{\~n}a for coding~\cite{saldana2021coding}. We consider these well-established references suitable for our case, where a classification of themes emerging from participant responses is   required. We did not adopt the more powerful Grounded Theory (GT) framework~\cite{hoda2021socio}, given that our problem is focused on already pre-defined, intuitive, macro-categories  (\textit{strategies}, \textit{benefits}, \textit{challenges}, \textit{differences}), and data collection is not intertwined with data analysis as in GT.  

\paragraph{Internal Validity}
the participants were involved in a course, and the overall assignment was part of their evaluation. This could have biased their activity, as the participants could feel compelled to produce more, and possibly unreliable, information. Furthermore, reflections were written \textit{after} performing the different tasks, thus leading to possible recall bias. These aspects could not be entirely mitigated. The qualitative analysis relies on data interpretation, which entails subjectivity. To mitigate this, we triangulated the produced codebooks between researchers, with open-coding activities followed by closed-coding ones on a different portion of the datasets, and with meetings to consolidate and homogenise the identified themes. We also share the raw and tagged data, the codebooks, including examples of participants' quotes for each theme identified, and we report the data analysis procedure with details~\cite{ferrari2022supplementary}, as  recommended for qualitative studies~\cite{hoda2021socio}.

\paragraph{External Validity}
this is an experimental simulation, and thus its results may not apply to real-world cases. However, this type of research strategy is rather close to a study in a natural setting~\cite{stol18}, thus reaching a reasonable compromise between realism and generalisability. The results apply to cases that are similar to ours, in which IBE is followed by ASE. Though we used students as participants, an accepted practice in software engineering~\cite{falessi2018empirical}, most of them are \textit{professionals}, and thus our results combine the viewpoint of both novice and advanced analysts. A residual threat is the usage of a single scenario to elicit our data, similarly to other studies in requirements engineering~\cite{debnath21,pitts2007improving}. Following case-based generalisation~\cite{wieringa2015six}, we deem the scenario as representative of social apps with different user profiles (e.g., managers, staff), multimedia-sharing features, and a specific application domain. This may have restricted the number of similar applications found by the participants during the experimental simulation. Different themes may be identified with other types of apps, e.g., more domain-generic ones. 
About the completeness of the findings, we performed a SLR to complement the codebooks. This applies to RQ2 and as a by-product  to RQ3, as data from RQ2 were used as input.

\paragraph{Literature Review} we followed the widely adopted guidelines from Kitchenham~\cite{kitchenham2004procedures}, and adopted the assumptions of Mart{\'\i}nez-Fern{\'a}ndez \textit{et al.}~\cite{martinez2022software} for the choice of Scopus as a single search engine. The search string is narrow, and we may have unavoidably missed relevant publications. We argue that this risk is acceptable, given that the SLR is a secondary source of information. Finally, while our research focuses on ASE, the SLR searched for studies about app store analysis. Though this is a broader activity, (1) we are not aware of studies specifically focused on benefits/challenges of ASE, and (2) in our data extraction we considered solely those benefits/challenges that are applicable to ASE.  





\section{Conclusion}
\label{sec:conclusion}
ASE is the practice of using app stores as a source of inspiration by selecting apps considered relevant for the product under development, understanding how they work, and browsing their reviews. While ASE is typically used in industry during the development process, the literature offers very little insight into this practice. 
In this work, we contribute to filling this gap by identifying the most commonly used strategies to select apps and the benefit and challenges that analysts associate with ASE. 
We conducted an experimental simulation with 58 analysts and performed a systematic analysis of the collected data. The results of this work contribute to the \textit{theory} of requirements engineering, providing the initial foundations of a well-known but yet not formalized activity, and outlining research and practice directions. 
In future works, we plan to: (1) survey practitioners to  quantify how relevant are the identified benefits and challenges from their viewpoint; (2)~evaluate the effect of specific inspiration strategies on the final requirements of a product.



\ifCLASSOPTIONcaptionsoff
  \newpage
\fi



\bibliographystyle{IEEEtran}
\bibliography{IEEEabrv,paper}
%


%




\end{document}